\documentclass[pre,twocolumn, superscriptaddress]{revtex4-1}

\usepackage{amsmath}
\usepackage{hyperref}

\usepackage{graphicx}
\usepackage{color}
\usepackage{Tomspeak}
\newcommand{\cross}{\times}
\newcommand{\Lop}{\mathsf{L \kern-4pt L}}
\newcommand{\Kop}{\mathsf{K \kern-6pt K}}
\newcommand{\hop}{\mathsf{h \kern-4pt h}}
\newcommand{\Eop}{\mathsf{E \kern-5pt E}}
\newcommand{\Oseen}  {{\rm O \kern -7pt O}} 
\newcommand{\sm}[1]{{\scriptscriptstyle \mathbf #1}}
\newcommand{\set}[1]{\{\vector{#1}_i\}} 
\newcommand{\comp}[1]{^{\scriptscriptstyle (#1)}}

\begin{document}
\title{Adapting the Teubner reciprocal relations for stokeslet objects}
\date{\today}
\author{Thomas A. Witten}
\email{t-witten@uchicago.edu}
\author{Aaron Mowitz}
\email{amowitz@uchicago.edu}
\affiliation{Department of Physics and James Franck Institute, University of Chicago, Chicago, Illinois 60637, United States.}

\begin{abstract}
Self-propelled colloidal swimmers  move by pushing the adjacent fluid backwards.  The resulting motion of an asymmetric body depends on  the profile of pushing velocity over its surface.  We describe a method of predicting the motion arising from arbitrary velocity profiles over a given body shape, using a discrete-source ``stokeslet" representation.  The net velocity and angular velocity is a sum of contributions from each point on the surface.  The contributions from a given point depend only on the shape.  We give a numerical method to find these contributions in terms of the stokeslet positions defining the shape. Each contribution is determined by linear operations on the Oseen interaction matrix between pairs of stokeslets.  We first adapt the Lorentz Reciprocal Theorem to discrete sources. We then use the theorem  to implement the method of Teubner\cite{Teubner:1982kq} to determine electrophoretic mobilities of nonuniformly charged bodies.

\end{abstract}
\maketitle

\section{Introduction}\label{sec:introduction}
Many forms of propulsion of microscopic objects through fluids work via {\em phoresis}: the surface of the object pushes the adjacent fluid to create a thin slip layer of nonzero relative velocity\cite{Moran:2017rm}.  Electrophoresis in an ionic fluid is a prime example; thermophoresis and chemophoresis behave analogously\cite{anderson1989colloid, Stone:1996fk, Eslahian:2014fj, Moyses:2016yq}.  In active propulsion chemical reactions at the surface create the propulsive surface flow\cite{Walther:2013jt, Maass:2016sy}.  Similarly many living organisms propel themselves via a layer of beating cilia on the surface\cite{Marchetti:2013pi}.  Even when the profile of the slip velocity is known, determining the resulting motion is challenging, especially when the object has an asymmetric shape.  Here we describe a general method of determining motion of an asymmetric self-propelling object.

We describe this method in the important context of  electrophoretic motion of a body of given shape and charge distribution\cite{anderson1989colloid,AndersonSpheres,AndersonEllipsoids,AjdariLong,CHAE:1995qd,Allison:2001db,Delgado:2007qy}.
On one level these motions amount to a mere linear response.  Yet the motions can be complex and unintuitive.  The body may rotate steadily with no electrophoretic translation; it may translate perpendicular to the applied field\cite{AjdariLong}. The body may move with an intrinsic chirality even though its shape is not chiral.  Analogous sedimentary motions were shown to enable coherent control over a dispersion of like objects\cite{Moths-Witten2}.  

In recent times the experimental importance of such objects has steadily risen.  Increasingly experiments study dispersions of colloidal particles of a common, asymmetric shape\cite{Han:2016cw, Sacanna:2011fk, Meng560}.  These bodies may be unevenly charged.  The ability to manipulate such bodies by harnessing their asymmetry is increasingly desirable.  Accordingly, there is an increasing need to predict the distinctive motions for a given shape and charge distribution.  Current methods of making such predictions are ungainly or limited in scope\cite{Teubner:1982kq, Delgado:2007qy}.  

In 1982 Teubner\cite{Teubner:1982kq, Burelbach:2019sz} made an insightful simplification of this problem.  He showed that the influence of charge distribution on a given body could be distilled into a set of six functions defined over its surface.  Each function amounted to the distribution of hydrodynamic shear stress over that surface when certain specified motions were imposed.  Using these six functions one may find the motion of the given body with arbitrary charge distributions.  His findings exploited the long-known Lorentz Reciprocal relations for stokes flow \cite{Happel-Brenner}.  However, finding these functions requires solving the Stokes equations for the fluid around the body in arbitrary motion.  Predicting the motion also required determining the applied electric field over the surface.  

One may avoid dealing with these equations by approximating the body as a ``stokeslet object".  
A stokeslet object is defined by a set of $N$ point sources or stokeslets $i$ applied at fixed relative positions $\vector r_i$ in a fluid.  Any  force $\vector f_i$ on a stokeslet causes a proportional flow around $\vector r_i$ given by a $3\cross 3$ response tensor $\Lop$ discussed below. A solid body can be approximated as a stokeslet object by replacing the body with a large number of stokeslets distributed over its surface\cite{Mowitz:2017kx}.  By giving the stokeslets electric charge as well as hydrodynamic forcing, this approximation was shown to predict electrophoretic mobility successfully in simple cases\cite{Mowitz:2017kx, Braverman:fq}.

This stokeslet approach raises a question: What is the analog of Teubner's insight for stokeslet objects?  That is, what is the analog of the six functions that allow treatment of arbitrary charge distributions on the object?  Here we answer this question and determine the analogous functions.  This approach provides a simple complement to the differential equations of the Teubner theory.  

The stokeslet approach also allows a simple understanding of the Lorentz Reciprocal Theorem \cite{Happel-Brenner}.  This theorem applies to flow fields in a region confined by defined boundaries of some given shape and differing in the stresses at these boundaries.  It follows from the multilinearity of the flow to the stress and the equality of external work and viscous dissipation.  These same features dictate the linear response of the velocity field to a point source of force, or stokeslet.  As in the Lorentz Reciprocal Theorem, the linear relation between the stokeslet forces and their velocity fields shows how two different velocity fields interact.  

First, we recall how stokeslets create velocity fields.  Next, we formulate the analog of the Lorentz Reciprocal relation\cite{Happel-Brenner} for stokeslet objects.  Finally, we apply this reciprocal relation to derive expressions for the force, torque and motion of an arbitrarily charged stokeslet object in an electric field.  
 
\section{superposition of stokeslet responses}\label{sec:superposition}

In the Stokes regime \cite{Happel-Brenner} of weak flow in a fluid at rest at infinity, the  velocity field of the fluid $\vector v(\vector r)$ is a linear function of the external forces acting on it.  A point force $\vector f_j$ acting at a position $\vector r_j$ produces a velocity field $\vector v_j(\vector r)$ proportional to $\vector f_j$.  That is, 
\begin{equation}\label{eq:Oseen}
\vector v_j(\vector r) = \Lop(\vector{\Delta r}) \cdot \vector f_j ,
\end{equation}
where $\Lop(\vector{\Delta r})$ is a $3\cross 3$ linear response tensor that depends on the displacement $\vector r - \vector r_j \definedas \vector{\Delta r}$ and the fluid's viscosity\footnote{The explicit form of $\Lop$ is well known and is called the Oseen tensor\cite{Happel-Brenner}.    However the present derivation doesn't depend on this explicit form.  One may readily verify that the Oseen tensor satisfies the symmetry property shown for $\Lop_{ij}$ below.
}.
We note that this $\vector v_j(\vector r)$ does not depend on any assumed motion of the forcing point $\vector r_j$ itself.  Sec. \ref{sec:stokeslet motion} discusses the physical motion leading to the $\vector f_j$.
\begin{figure}[htbp]
\includegraphics[width=\columnwidth]{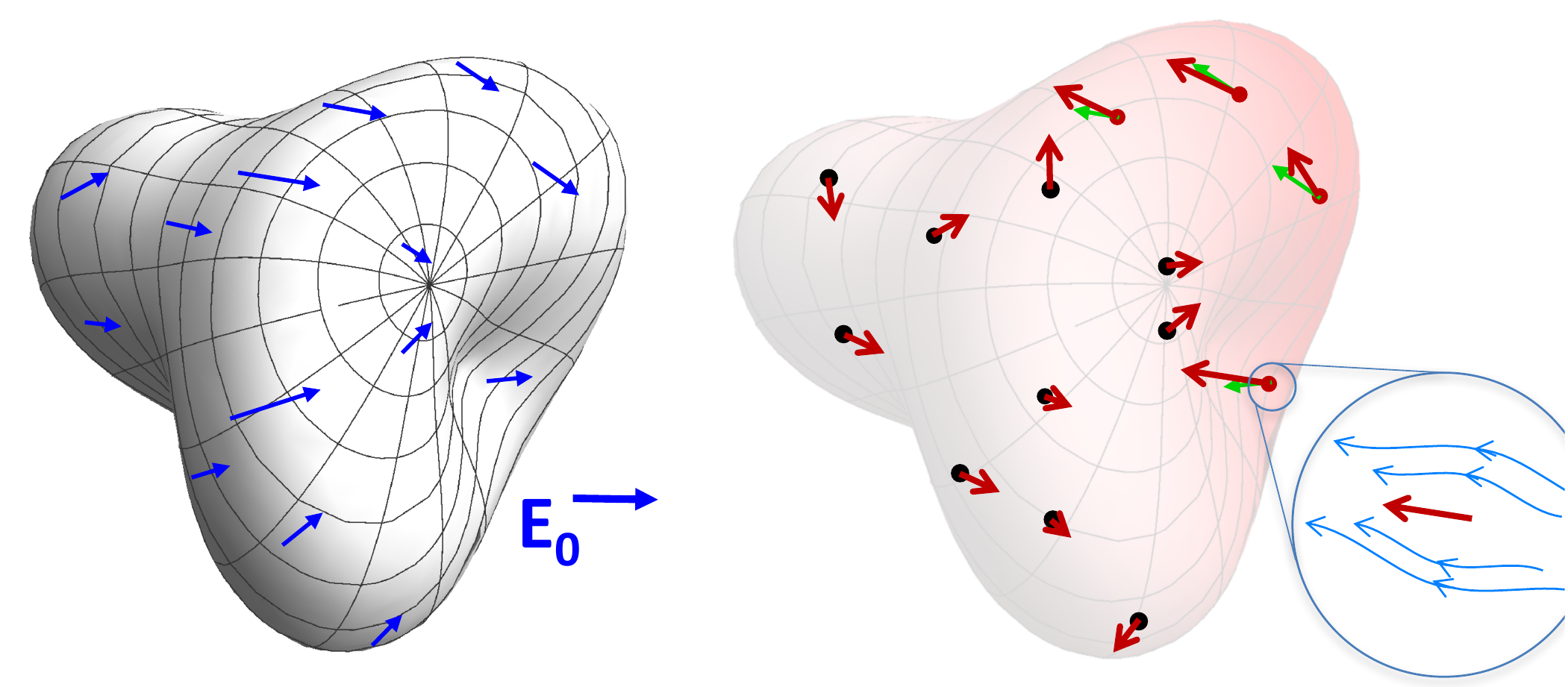}
\caption{\label{fig:cartoon}Electrophoresis of a partially charged stokeslet object. Left: object in an external field $\vector E_0$ pointing to the right.  Colored arrows show tangential surface field $\vector E\sm{S}$, depending only on shape.  Right: stokeslet representation of the object.  Large dots show the stokeslet positions.  Colored dots represent charged regions. Green, light arrows show slip velocity at the colored dots, proportional to $\vector E\sm{S}$ and to the local zeta potential $\zeta_i$. Red, heavy arrows show stokeslet forces that together produce matching velocities at the colored dots and zero velocity at the black dots. They also generally create flow in the interior. This flow is not relevant for electrophoretic motion. Inset at right: colored streamlines indicate the Oseen velocity field $\Lop(r) \cdot\vector f_i$ around one stokeslet.}

\end{figure}

We consider a given set of $N$ stokeslets at positions $\vector r_1, \vector r_2, ..., \vector r_N = \set{r}$.  If forces $\set{f}$ are applied to these stokeslets, a velocity field $\vector v(\vector r)$ results, as shown in Fig.\ref{fig:cartoon}.    In particular, the velocity of the fluid at stokeslet $i$ is the sum of contributions $\vector v_{ij}$ due to the $\vector f_j$ via Eq.  \eqref{eq:Oseen}.  
\begin{equation}\label{eq:fj2vij}
\vector v_{ij} = \Lop_{ij} \cdot \vector f_j, 
\end{equation}
where $\Lop_{ij} \definedas \Lop(\vector r_i - \vector r_j)$.
The $\set{v}$ are then determined by superposition: 
\begin{equation} \label{eq:fj2vi}
\vector v_i = \sum_{j=1}^N \vector v_{ij} = \sum_{j=1}^N \Lop_{ij} \cdot \vector f_j
\end{equation} 
One may determine the $\set{f}$ that generate a desired $\set{v}$ by solving the $N$ simultaneous linear equations of Eq. \eqref{eq:fj2vi}\footnote{
The diagonal elements $\Lop_{ii}$ are not defined for point stokeslets.  A better approximation for the velocity at $\vector r_i$ owing to the force at site $i$ is required.  It is convenient to replace point the point force by distributing it uniformly over over some small region representing the continuum force on the fluid near site $i$.  This procedure is not unique, but its effect on determining the $\vector f_i$ diminishes as the number of stokeslets increases.  The validity of the reciprocal relation of Sec. \ref{sec:reciprocity} relies only on using the same regularization and hence the same $\Lop$ for determining both velocity fields considered.
}
\footnote{
A similar procedure treats motion caused by an external force.  Here one requires that all the stokeslets move at a common velocity $\vector V$ and that the stokeslets provide sufficient screening that the interior of the object also moves at velocity $\vector V$.  Then the total of the stokeslet forces is the force required for this motion\cite{Mowitz:2017kx}}.

Together, the $\set{v}$ and $\set{f}$ determine the rate $\dot W$ at which the stokeslet forces do work on the fluid. This work is necessarily quadratic in the $\set f$  When the stokeslets are placed together in the fluid, extra work is required for each stokeslet $i$ due to the velocity $v_i$ there due to the other stokeslets. The power $\dot w_{ij}$ associated with $\vector f_i$ and $\vector f_j$ is $\vector f_i \cdot \vector v_{ji} + \vector f_j \cdot \vector v_{ij}$ or 
\begin{equation}\label{eq:wij}
\vector f_i \cdot \Lop_{ij}\cdot  \vector f_j + \vector f_j \cdot\Lop_{ji} \cdot \vector f_i 
= \vector f_i\cdot (\Lop_{ij} + \Lop_{ji}^T) \cdot\vector f_j.
\end{equation}
Here $\Lop_{ij}^T$ denotes the transpose of the 3$\cross$3 matrix $\Lop_{ij}$.  

This expression implies that the response tensor $\Lop$ itself is symmetric; that is, $\Lop_{ji}^T = \Lop_{ij}$.  To verify this, we write  $\Lop_{ij}$ as the sum of its symmetric and antisymmetric parts, $\Lop_{ij}^S \definedas \half ( \Lop_{ij}+\Lop_{ji}^T)$ and $\Lop_{ij}^A \definedas \half ( \Lop_{ij}-\Lop_{ji}^T)$.  Eq. \eqref{eq:wij} shows that the $\dot w_{ij}$ depend on only the symmetric part $\Lop^S$, while the velocities $\vector v_{ij}$ are  the sum of contributions from both $\Lop^S$ and $\Lop^A$.   Thus if $\Lop^A$ were nonzero, it would give rise to a nonzero $\set{v}$ whose $\dot w_{ij} = 0$.  Now, no such flow can exist, since every flow that vanishes at infinity necessarily has a nonzero shear rate and thence a positive $\dot W$ from viscous dissipation.  Thus $\Lop^A$ vanishes and $\Lop$ is symmetric, as claimed.  

\section{reciprocity}\label{sec:reciprocity}

The symmetry of the response matrix $\Lop$ entails a reciprocal relation between different forces $\set{f}$ and the velocities $\set{v}$ they generate.  This is the stokeslet analog of the Lorentz Reciprocal Relation\cite{Happel-Brenner}.

We now suppose that a given set of stokeslets experiences two sets of forces: $\set{f}$ and $\set{g}$.  The $\set{f}$ generate a set of velocities $\set{v}$ as told above.  Likewise, the $\set{g}$ generate a different set of velocities denoted $\set{u}$.  Evidently part of the power $\dot W$ is the contribution from the forces $\set{g}$ acting on the velocities $\set{v}$: \ie $\sum_i \vector g_i \cdot \vector v_i$. Another part is the contribution from the $\set{f}$'s acting on the $\set{u}$'s.  Owing to the symmetry of $\Lop$, these two works can be readily seen to be equal:  both are equal to a symmetric sum over $\set{f}$ and $\set{g}$:
\begin{eqnarray}\label{eq:reciprocal}
\sum_{i=1}^N \vector g_i \cdot \vector v_i =& \sum_{i,j=1}^N \vector g_i \cdot \Lop_{ij} \cdot \vector f_j
= \sum_{i,j=1}^N \vector f_j \cdot \Lop^T_{ij} \cdot \vector g_i \nonumber\\
=& \sum_{i,j=1}^N \vector f_j \cdot \Lop_{ji} \cdot \vector g_i = \sum_{j=1}^N \vector f_j \cdot \vector u_j
. 
\end{eqnarray}
Thus the work done by $\set{f}$ on the $\set{g}$ velocities is equal to the work done by $\set{g}$ on the $\set{f}$ velocities.

\section{electrophoretic forces}\label{sec:forces}
 
We may use this reciprocal relationship to determine the forces on a charged colloidal body in a static electric field $\vector E_0$.   A body that is held stationary experiences a constraint force $\vector F$ and torque $\vector \tau$ proportional to $\vector E_0$.  To lowest order in $E_0$ the force and torque are given by two tensors $\Kop_F$ and $\Kop_\tau$ defined by
\begin{equation}\label{eq:Kops}
\vector F = \Kop_F\cdot \vector E_0 \quad \hbox{and} \quad
\vector \tau  = \Kop_\tau \cdot \vector E_0
.\end{equation}

We consider the typical case in which ions in the solvent strongly screen any electric field due to the body charges $q_i$.  Thus the field from the body charge is confined to a screening layer much thinner than the body's size\cite{AndersonSpheres}.  When the external field $\vector E_0$ is applied, the field near the insulating object is distorted, leaving a tangential surface field $\vector E\sm{S}(\vector r)$ at the surface induced by $\vector E_0$. This $\vector E\sm{S}$ is caused by $\vector E_0$ alone, and it depends only on the body shape and the $\vector E_0$.  The force and torque arise from a thin sheath\cite{AndersonSpheres} of moving screening charge around the charged portions of the object.  This sheath flow amounts to a local slip velocity field $\vector v_s(\vector r)$ proportional to the local charge density and to the surface electric field there.  The charge density is generally expressed in terms of the ``zeta potential" $\zeta(\vector r)$\cite{AndersonSpheres, Delgado:2007qy}\footnote{
The zeta potential that determines the slip velocity is the potential of the object surface relative to the bulk fluid. It depends on the local ionic environment of the surface in equilibrium without applied field $\vector E_0$.  It also depends on the local charge density on the surface and is proportional to this density for weakly charged regions.  Typical colloids in aqueous solvents have $\zeta$ in the range of tens of millivolts.  It is measured for a given type of surface by electrophoresis on a uniformly charged body with that type of surface.
}.  
In the corresponding stokeslet object, the slip velocity $\vector v_i$ at stokeslet $i$ may be expressed as  $\kappa ~\zeta_i$ times the surface electric field at stokeslet $i$, where $\kappa$ is a material constant of the solvent.\cite{AndersonSpheres}.  Thus the surface velocities $\vector v_i = \kappa~ \vector E\sm{S}_i~ \zeta_i$ may be considered known.  

However, the quantities needed to determine the object's eventual motion are not these $\set{v}$ but the force and torque $\vector F$ and $\vector \tau$.  These are necessarily some linear function of the $\set{v}$, which in turn depend on the $\{\zeta_i\}$.  Thus \eg the total force $\vector F$ in the 1 direction, $F\comp1$, must be of the form $\sum_{i=1}^N \vector h\sm{F1}_i \cdot \vector v_i$ for some set of coefficients $\vector h\sm{F1}_i$. This $\vector h\sm{F1}_i$ is evidently the green's function giving the $F\comp1$ response to the velocity $\vector v_i$. Using the reciprocity property of $\dot W$, we can determine this green's function, as we now show.  

The force component in the 1 direction $F\comp1$ is evidently the sum of its stokeslet contributions: $F\comp1 = \sum_{i=1}^N f\comp1_i$.  We wish to express this quantity in terms of the known $\set v$.  The reciprocity relation \eqref{eq:reciprocal} states that for any other set of forces $\vector g_i$ and their corresponding velocities $\set u$, the sum 
$\sum_{i=1}^N \vector f_i \cdot \vector u_i = \sum_{i=1}^N \vector g_i \cdot \vector v_i$ . In order to make the summand on the left become $f\comp1_i$, it suffices for $\vector u_i$ to be $(1, 0, 0)$ for all $i$.  The forces $\set{g}$ corresponding to these $\set{u}$ by definition satisfy 
\begin{equation}\label{eq:F1}
(1, 0, 0) = \sum_{j=1}^N \Lop_{ij}\cdot \vector g_j
\end{equation}
for all stokeslets $i$.  These $\set g$ depend only on the positions of the stokeslets; they are unrelated to the charges $\set q$ or their potentials $\set{\zeta}$.  Using these $\set g$, we conclude that
\begin{equation}
F\comp1 = \sum_{i=1}^N \vector g_i \cdot \vector v_i
\end{equation}

The $\set g$ obtained by solving Eq. \eqref{eq:F1} are thus the desired green's function $\set{h\sm{F1}}$.  The other green's functions $\set{h\sm{F2}}$ and $\set{h\sm{F3}}$ are found analogously by changing $(1, 0, 0)$ in Eq. \eqref{eq:F1} to $(0, 1, 0)$ and $(0, 0, 1)$.
Combining these results, the vector $\vector F$ may be be written in matrix form
\begin{equation}\label{eq:FOfv}
\vector F = \sum_{i=1}^N \hop\sm{F}_i \cdot \vector v_i ,
\end{equation}
where \eg the $1, 2$ matrix element of the $3\cross 3$ matrix $\hop\sm{F}_i$ is $h\sm{F1}_i\comp2$.  We denote $\hop\sm{F}$ as the force green's function.
 
In the same way we may find the 1-component of the total torque, $\tau\comp1$.  As before we use Eq. \eqref{eq:reciprocal}: we must find a set of velocities $\set{u}$ such that $\tau\comp1 = \sum_{i=1}^N \vector f_i\cdot \vector u_i$.  Indeed, $\vector \tau$ {\em is} the sum of contributions $\vector \tau_i$ from individual stokeslets, where $\vector \tau_i = \vector r_i \cross \vector f_i$. In particular $\tau\comp1_i = f\comp3_i r\comp2_i - f\comp2_i r\comp3_i = \vector f_i \cdot (0, -r\comp3_i, r\comp2_i )$.
Thus the needed velocities $\vector u_i$ are $\vector u_i = (0, -r\comp3_i, r\comp2_i )$.  Their corresponding forces $\vector h\sm{\tau1}$ are the solutions to the $3N$ equations 
\begin{equation}
(0, -r\comp3_i, r\comp2_i ) = \sum_{j=1}^N \Lop_{ij}\cdot \vector h\sm{\tau1}_j
\end{equation}
analogous to Eq. \eqref{eq:F1}.  Then using the reciprocal equation \eqref{eq:reciprocal}, 
\begin{equation}\label{eq:taucomp}
\tau\comp1 = \sum_{i=1}^N \vector h\sm{\tau1}_i \cdot\vector v_i
\end{equation} 
Analogous procedures give the coefficients $\set {h\sm{\tau2}}$ for $\tau\comp2$ and $\set{h\sm{\tau3}}$ for $\tau\comp3$.  Combining these into matrix form, we write the vector $\vector \tau$ in the form
\begin{equation}\label{eq:TauOfv}
\vector \tau = \sum_{i=1}^N \hop\sm{\tau}_i \cdot \vector v_i
\end{equation}
where the matrix element $[\hop\sm{\tau}_i]_{\alpha \beta}$ is the contribution of $v_i\comp\beta$ to $\tau\comp\alpha$.  This $\hop\sm{\tau}$ is the torque green's function analogous to $\hop\sm{F}$ for force.  

The two green's functions $\hop\sm{F}$ and $\hop\sm{\tau}$ allow us to find the response matrices $\Kop_F$ and $\Kop_\tau$ of Eq. \eqref{eq:Kops}.  
To this end we must determine the slip velocities $\set{v}$ in Eqs. \eqref{eq:TauOfv} and \eqref{eq:FOfv} for given surface fields $\set{E\sm{S}}$.
Each field $\vector E\sm{S}_i$ is in turn proportional to the external field $\vector E_0$.  In matrix notation the surface field at stokeslet $i$, is given by 
\begin{equation}
\vector E\sm{S}_i = \Eop_i\cdot \vector E_0 ,
\end{equation}
where \eg the 3x3 matrix element $[\Eop\sm{s}_i]_{1, 2}$ is the contribution to $E\sm{S}_i\comp1$ from $\vector E_0\comp2$.   As explained above, each $\vector E\sm{S}_i$ produces a velocity $\vector v_i = \kappa \zeta_i~ \vector E\sm{S}_i$.  Using this expression for $\vector v_i$ in Eq. \eqref{eq:F1} we obtain for \eg $F\comp1$ when $E_0 = (0,1,0)$
\begin{equation}
F\comp1 = \kappa\sum_{i=1}^N \vector h\sm{F1}_i\cdot \Eop_i\cdot (0,1,0) ~\zeta_i.  
\end{equation}
The full vector $\vector F$ then has the same form with $\vector h$ replaced by $\hop$:
\begin{equation}
\vector F = \left (\kappa \sum_{i=1}^N \zeta_i~ \hop\sm{F}_i \cdot \Eop_i\right ) \cdot\vector E_0
\end{equation}

The expression in $(...)$ is evidently the $\Kop_F$ response matrix of Eq. \eqref{eq:Kops}.  The analogous equations for $\vector \tau$ give the $\Kop_\tau$ response matrix.

\section{electrophoretic motion}\label{sec:motion}
The above procedure gives the force and torque exerted on the fluid by the immobilized object with a particular orientation in the $\vector E_0$ field.  The same force and torque must be externally applied to the object in order to hold it in place.  If the object is now freed from its constraints, it will translate and rotate.  This translation and rotation do not alter the electrophoretic force and torque calculated above; these are determined by the viscous drag across the slip layer, and they depend only on the relative velocity between the local surface and the adjacent screening charge\footnote{
For general stokeslet objects, this locality may not be well defined, since an arbitrary set of stokeslets need not resemble any smooth surface.  However, if stokeslets are arranged over a smooth surface with spacing much smaller than the local inverse curvature, the stokeslet object may approximate the corresponding smooth body, as noted above.  Then the above reasoning applies, a stokes mobility tensor may be determined, and the $\vector V$ and $\vector \Omega$ may be calculated\protect{\cite{Mowitz:2017kx}}
}.  
Without constraint forces, these electric forces are balanced by drag forces due to the motion.  These drag forces themselves are related to the $6 \cross 6$ Stokes mobility tensor ${\mathbf M}$ relating the velocity $\vector V$ and angular velocity $\vector \Omega$ to $\vector F$ and $\vector \tau$\cite{Happel-Brenner} via 
\begin{equation} \label{eq:stokesMobility}
(\vector V, \vector \Omega) \definedas {\mathbf M} \cdot (\vector F, \vector \tau)
\end{equation} 
Calculation of the stokes drag tensor ${\mathbf M}$ implicitly gives the set of stokeslet forces $\set{\ell}$\cite{Mowitz:2017kx} for a given velocity and angular velocity, \eg $\vector V = (1, 0, 0)$ and $\vector \Omega = 0$.  

\section{discussion}\label{sec:discussion}
Here we discuss several questions about the relation of our approach to prior approaches and the generalization of our methods to other forms of driven motion.

\subsection{Flow and dissipation} \label{sec:Dissipation}
In the electrophoresis of a solid object, the motion produces a shear flow outside the object.  In addition there is a strong shear flow across the thin electrostatic slip layer.  This latter flow generates arbitrarily more dissipation than does the exterior flow.  Reducing the thickness of this layer by a factor $\lambda$ while increasing the charge density by the same factor has no effect on the zeta potential or the motion but increases the viscous dissipation by a factor $\lambda$\cite{anderson1989colloid}.  

One may ask how these dissipations are related to the power $\dot W$ discussed above.  There the surface velocity was generated by a single layer of stokeslets, not by a thin shear layer.  This means that the boundary condition that the fluid within the surface remain stationary was not imposed.  As we have seen above, this solid-surface boundary condition, though physical, is not relevant for the electrophoretic force (as is implicit in Anderson's work\cite{anderson1989colloid}).  Without this boundary condition, the interior is free to flow.  In our forthcoming explicit calculations\cite{Braverman:fq}, the $\set{f}$ generate a circulating flow inside the object surface as well as an exterior flow.  Still, the force and torque agree well with known electrophoresis results.  Since there is no thin shear layer, the dissipation $\dot W$ is far smaller than that generated in a real solid object.

Exterior flow from electrophoresis is well known to be qualitatively weaker than from sedimentation at the same velocity.  Sedimentation generates a force-monopole flow at infinity falling off inversely with distance\footnote{
The functional form of this falloff is the well-known Oseen tensor mentioned above.}, 
while electrophoretic flow falls off inversely with distance squared or faster.  In our formulation the object is held fixed; the electrically produced force and torque are balanced by a constraining  force and torque.  The resulting flow has a force monopole, as with sedimentation.  However, when the constraint is released, Stokes drag generated by the motion compensates exactly for this constraint force. There is then no external force and the force monopole part of the external flow vanishes.  The external torque also vanishes; this further constrains the long-distance falloff of the velocity field.  We note that calculating the Stokes drag for a given motion {\em does} require the boundary condition of a solid surface. This creates a major difference between the stokeslet forces $\set{\ell}$ defined below Eq. \eqref{eq:stokesMobility} for sedimentation and the forces $\set{h}$ of Eqs. \eqref{eq:FOfv} and \eqref{eq:taucomp} for electrophoresis.    

\subsection{Sufficiency of the green's functions \protect $\hop$ }\label{sec:sufficiency}
Teubner's method adapted here provides the complete information to determine the motion of the object with an arbitrary surface velocity field $\set{v}$, using the two green's function matrices $\hop\sm{F}$ and $\hop\sm{\tau}$.  Thus in general there are many possible $\set{v}$'s on the object that produce the same motion $\vector V$ and $\vector \Omega$.  In particular, there is a sedimentation force and torque that produces this $\vector V$ and $\vector \Omega$.  One may ask whether the many $\set{v}$ that give the same $\vector V$ and $\vector \Omega$ behave equivalently or have a common set of stokeslet forces $\set{f}$.  In fact they are not equivalent.  Each has a distinctive flow in its vicinity, though the $\vector V$ and $\vector \Omega$ depend only on particular moments of this flow.  However, other hydrodynamic behaviors in general depend on different features of $\set{v}$.  For example, the hydrodynamic interaction of two such objects is sensitive to these other features\cite{TomerHaim2,TomerHaim1}.  

\subsection{Motion of the stokeslets}  \label{sec:stokeslet motion} 
In the foregoing we have discussed how stokeslet forces create a desired set of surface velocities.  It may seem odd that the motion of the stokeslets themselves did not appear.  The actual source of this force is the layer of screening ions near the surface.  However, when we represent this force by stokeslets, we need not consider this motion explicitly.  If we consider the forces to be coming from small spheres exerting drag forces on the adjacent fluid, their motion for given $\set{f}$ depends on their drag coefficients, which in turn depends on their size. If one were to choose very small stokeslets, their speed would be arbitrarily large, even with fixed surface velocities $\vector v_s(\vector r)$.  Their size influences the total work done on the fluid, but not the continuum motion of the fluid of importance here.  The situation is different when stokeslet objects are used to determine sedimentation forces. Then the stokeslets move with the object\cite{Kirkwood-Riseman}, and their drag coefficients are important.  

\subsection{Driven and active colloids} \label{sec:Driven}
Though our discussion has been framed in terms of charged objects in an electric field, it is applicable more broadly.  The essential mechanism of electrophoresis is motion driven by an imposed slip velocity field $\vector v_i$ over a given surface.  As noted above, an imposed slip velocity can be driven by many sources other than ion flow.  Indeed, any nonuniform potential that affects the energy of the surface can drive similar phoretic slip velocities.  Important examples are chemical potentials from concentration or temperature gradients\cite{anderson1989colloid, Stone:1996fk, Eslahian:2014fj, Moyses:2016yq}.  Moreover, these gradients can be generated by chemical reactions in the colloidal object itself, resulting in active particle motion\cite{Walther:2013jt, Maass:2016sy}.  Finally, the slip velocities may be generated by mechanically driven actuators on the surface.  Many living organisms move by generating a beating motion of cilia on their surface\cite{Marchetti:2013pi}.    

In many of these active systems the origin of the slip velocity field is different than in electrophoresis.  On the one hand, the velocity field $\vector v_s(\vector r)$ may be fixed with respect to the body, with no dependence on an external vector such as $\vector E_0$.  On the other hand $\vector v_s(\vector r)$ may not be determined by the structure of the object.  The slip velocity may for example require triggering by some symmetry-breaking initial motion\cite{Zottl:2014ye}.  The methods above do not address the cause of the slip velocity; they only predict its consequences in generating force and motion.  Still, these methods appear useful in understanding an important aspect of these active motions.

\section{conclusion}\label{sec:conclusion}
Here we have extended the Teubner methodology from continuum shapes to simplified discrete objects called stokeslet objects.  We have shown its formal validity, but not its practical utility.  This will only be established when the method uncovers novel behavior of driven colloids that can be demonstrated experimentally.  Our paper in preparation\cite{Braverman:fq} gives strong evidence that stokeslet objects are useful for predicting electrophoresis for a wide range of shapes and charge distributions.  Thus there is reason for optimism that the reciprocal procedure developed here will prove useful in practice.  Meanwhile, the reasoning used here may shed light on the conceptual basis of the Lorentz Reciprocity Theorem.

\acknowledgements
\label{sec:acknowledgements}
The authors are grateful to Naomi Oppenheimer and Tomer Goldfriend for important critiques on an earlier manuscript.  This work was partially supported by the University of Chicago Materials Research Science and Engineering Center, which is funded by the National Science Foundation under award number DMR-1420709.

\bibliography{stokesletElectrophoresis}

\begin{thebibliography}{33}%
\makeatletter
\providecommand \@ifxundefined [1]{%
 \@ifx{#1\undefined}
}%
\providecommand \@ifnum [1]{%
 \ifnum #1\expandafter \@firstoftwo
 \else \expandafter \@secondoftwo
 \fi
}%
\providecommand \@ifx [1]{%
 \ifx #1\expandafter \@firstoftwo
 \else \expandafter \@secondoftwo
 \fi
}%
\providecommand \natexlab [1]{#1}%
\providecommand \enquote  [1]{``#1''}%
\providecommand \bibnamefont  [1]{#1}%
\providecommand \bibfnamefont [1]{#1}%
\providecommand \citenamefont [1]{#1}%
\providecommand \href@noop [0]{\@secondoftwo}%
\providecommand \href [0]{\begingroup \@sanitize@url \@href}%
\providecommand \@href[1]{\@@startlink{#1}\@@href}%
\providecommand \@@href[1]{\endgroup#1\@@endlink}%
\providecommand \@sanitize@url [0]{\catcode `\\12\catcode `\$12\catcode
  `\&12\catcode `\#12\catcode `\^12\catcode `\_12\catcode `\%12\relax}%
\providecommand \@@startlink[1]{}%
\providecommand \@@endlink[0]{}%
\providecommand \url  [0]{\begingroup\@sanitize@url \@url }%
\providecommand \@url [1]{\endgroup\@href {#1}{\urlprefix }}%
\providecommand \urlprefix  [0]{URL }%
\providecommand \Eprint [0]{\href }%
\providecommand \doibase [0]{http://dx.doi.org/}%
\providecommand \selectlanguage [0]{\@gobble}%
\providecommand \bibinfo  [0]{\@secondoftwo}%
\providecommand \bibfield  [0]{\@secondoftwo}%
\providecommand \translation [1]{[#1]}%
\providecommand \BibitemOpen [0]{}%
\providecommand \bibitemStop [0]{}%
\providecommand \bibitemNoStop [0]{.\EOS\space}%
\providecommand \EOS [0]{\spacefactor3000\relax}%
\providecommand \BibitemShut  [1]{\csname bibitem#1\endcsname}%
\let\auto@bib@innerbib\@empty
\bibitem [{\citenamefont {Teubner}(1982)}]{Teubner:1982kq}%
  \BibitemOpen
  \bibfield  {author} {\bibinfo {author} {\bibfnamefont {M.}~\bibnamefont
  {Teubner}},\ }\href
  {http://scitation.aip.org/content/aip/journal/jcp/76/11/10.1063/1.442861}
  {\bibfield  {journal} {\bibinfo  {journal} {The Journal of Chemical Physics}\
  }\textbf {\bibinfo {volume} {76}},\ \bibinfo {pages} {5564} (\bibinfo {year}
  {1982})}\BibitemShut {NoStop}%
\bibitem [{\citenamefont {Moran}\ and\ \citenamefont
  {Posner}(2017)}]{Moran:2017rm}%
  \BibitemOpen
  \bibfield  {author} {\bibinfo {author} {\bibfnamefont {J.~L.}\ \bibnamefont
  {Moran}}\ and\ \bibinfo {author} {\bibfnamefont {J.~D.}\ \bibnamefont
  {Posner}},\ }\href {\doibase 10.1146/annurev-fluid-122414-034456} {\bibfield
  {journal} {\bibinfo  {journal} {Annual Review of Fluid Mechanics}\ }\textbf
  {\bibinfo {volume} {49}},\ \bibinfo {pages} {511} (\bibinfo {year} {2017})},\
  \Eprint
  {http://arxiv.org/abs/https://doi.org/10.1146/annurev-fluid-122414-034456}
  {https://doi.org/10.1146/annurev-fluid-122414-034456} \BibitemShut {NoStop}%
\bibitem [{\citenamefont {Anderson}(1989)}]{anderson1989colloid}%
  \BibitemOpen
  \bibfield  {author} {\bibinfo {author} {\bibfnamefont {J.~L.}\ \bibnamefont
  {Anderson}},\ }\href@noop {} {\bibfield  {journal} {\bibinfo  {journal}
  {Annual review of fluid mechanics}\ }\textbf {\bibinfo {volume} {21}},\
  \bibinfo {pages} {61} (\bibinfo {year} {1989})}\BibitemShut {NoStop}%
\bibitem [{\citenamefont {Stone}\ and\ \citenamefont
  {Samuel}(1996)}]{Stone:1996fk}%
  \BibitemOpen
  \bibfield  {author} {\bibinfo {author} {\bibfnamefont {H.~A.}\ \bibnamefont
  {Stone}}\ and\ \bibinfo {author} {\bibfnamefont {A.~D.}\ \bibnamefont
  {Samuel}},\ }\href@noop {} {\bibfield  {journal} {\bibinfo  {journal}
  {Physical review letters}\ }\textbf {\bibinfo {volume} {77}},\ \bibinfo
  {pages} {4102} (\bibinfo {year} {1996})}\BibitemShut {NoStop}%
\bibitem [{\citenamefont {Eslahian}\ \emph {et~al.}(2014)\citenamefont
  {Eslahian}, \citenamefont {Majee}, \citenamefont {Maskos},\ and\
  \citenamefont {W{\"u}rger}}]{Eslahian:2014fj}%
  \BibitemOpen
  \bibfield  {author} {\bibinfo {author} {\bibfnamefont {K.~A.}\ \bibnamefont
  {Eslahian}}, \bibinfo {author} {\bibfnamefont {A.}~\bibnamefont {Majee}},
  \bibinfo {author} {\bibfnamefont {M.}~\bibnamefont {Maskos}}, \ and\ \bibinfo
  {author} {\bibfnamefont {A.}~\bibnamefont {W{\"u}rger}},\ }\href@noop {}
  {\bibfield  {journal} {\bibinfo  {journal} {Soft Matter}\ }\textbf {\bibinfo
  {volume} {10}},\ \bibinfo {pages} {1931} (\bibinfo {year}
  {2014})}\BibitemShut {NoStop}%
\bibitem [{\citenamefont {Moyses}\ \emph {et~al.}(2016)\citenamefont {Moyses},
  \citenamefont {Palacci}, \citenamefont {Sacanna},\ and\ \citenamefont
  {Grier}}]{Moyses:2016yq}%
  \BibitemOpen
  \bibfield  {author} {\bibinfo {author} {\bibfnamefont {H.}~\bibnamefont
  {Moyses}}, \bibinfo {author} {\bibfnamefont {J.}~\bibnamefont {Palacci}},
  \bibinfo {author} {\bibfnamefont {S.}~\bibnamefont {Sacanna}}, \ and\
  \bibinfo {author} {\bibfnamefont {D.~G.}\ \bibnamefont {Grier}},\ }\href@noop
  {} {\bibfield  {journal} {\bibinfo  {journal} {Soft Matter}\ }\textbf
  {\bibinfo {volume} {12}},\ \bibinfo {pages} {6357} (\bibinfo {year}
  {2016})}\BibitemShut {NoStop}%
\bibitem [{\citenamefont {Walther}\ and\ \citenamefont
  {M{\"u}ller}(2013)}]{Walther:2013jt}%
  \BibitemOpen
  \bibfield  {author} {\bibinfo {author} {\bibfnamefont {A.}~\bibnamefont
  {Walther}}\ and\ \bibinfo {author} {\bibfnamefont {A.~H.~E.}\ \bibnamefont
  {M{\"u}ller}},\ }\bibfield  {booktitle} {\emph {\bibinfo {booktitle}
  {Chemical Reviews}},\ }\href {\doibase 10.1021/cr300089t} {\bibfield
  {journal} {\bibinfo  {journal} {Chemical Reviews}\ }\textbf {\bibinfo
  {volume} {113}},\ \bibinfo {pages} {5194} (\bibinfo {year}
  {2013})}\BibitemShut {NoStop}%
\bibitem [{\citenamefont {Maass}\ \emph {et~al.}(2016)\citenamefont {Maass},
  \citenamefont {Kr{\"u}ger}, \citenamefont {Herminghaus},\ and\ \citenamefont
  {Bahr}}]{Maass:2016sy}%
  \BibitemOpen
  \bibfield  {author} {\bibinfo {author} {\bibfnamefont {C.~C.}\ \bibnamefont
  {Maass}}, \bibinfo {author} {\bibfnamefont {C.}~\bibnamefont {Kr{\"u}ger}},
  \bibinfo {author} {\bibfnamefont {S.}~\bibnamefont {Herminghaus}}, \ and\
  \bibinfo {author} {\bibfnamefont {C.}~\bibnamefont {Bahr}},\ }\href {\doibase
  10.1146/annurev-conmatphys-031115-011517} {\bibfield  {journal} {\bibinfo
  {journal} {Annual Review of Condensed Matter Physics}\ }\textbf {\bibinfo
  {volume} {7}},\ \bibinfo {pages} {171} (\bibinfo {year} {2016})}\BibitemShut
  {NoStop}%
\bibitem [{\citenamefont {Marchetti}\ \emph {et~al.}(2013)\citenamefont
  {Marchetti}, \citenamefont {Joanny}, \citenamefont {Ramaswamy}, \citenamefont
  {Liverpool}, \citenamefont {Prost}, \citenamefont {Rao},\ and\ \citenamefont
  {Simha}}]{Marchetti:2013pi}%
  \BibitemOpen
  \bibfield  {author} {\bibinfo {author} {\bibfnamefont {M.~C.}\ \bibnamefont
  {Marchetti}}, \bibinfo {author} {\bibfnamefont {J.~F.}\ \bibnamefont
  {Joanny}}, \bibinfo {author} {\bibfnamefont {S.}~\bibnamefont {Ramaswamy}},
  \bibinfo {author} {\bibfnamefont {T.~B.}\ \bibnamefont {Liverpool}}, \bibinfo
  {author} {\bibfnamefont {J.}~\bibnamefont {Prost}}, \bibinfo {author}
  {\bibfnamefont {M.}~\bibnamefont {Rao}}, \ and\ \bibinfo {author}
  {\bibfnamefont {R.~A.}\ \bibnamefont {Simha}},\ }\href {\doibase
  10.1103/RevModPhys.85.1143} {\bibfield  {journal} {\bibinfo  {journal} {Rev.
  Mod. Phys.}\ }\textbf {\bibinfo {volume} {85}},\ \bibinfo {pages} {UNSP 1143}
  (\bibinfo {year} {2013})}\BibitemShut {NoStop}%
\bibitem [{\citenamefont {Anderson}(1985)}]{AndersonSpheres}%
  \BibitemOpen
  \bibfield  {author} {\bibinfo {author} {\bibfnamefont {J.~L.}\ \bibnamefont
  {Anderson}},\ }\href {\doibase
  http://dx.doi.org/10.1016/0021-9797(85)90345-5} {\bibfield  {journal}
  {\bibinfo  {journal} {Journal of Colloid and Interface Science}\ }\textbf
  {\bibinfo {volume} {105}},\ \bibinfo {pages} {45 } (\bibinfo {year}
  {1985})}\BibitemShut {NoStop}%
\bibitem [{\citenamefont {Fair}\ and\ \citenamefont
  {Anderson}(1989)}]{AndersonEllipsoids}%
  \BibitemOpen
  \bibfield  {author} {\bibinfo {author} {\bibfnamefont {M.}~\bibnamefont
  {Fair}}\ and\ \bibinfo {author} {\bibfnamefont {J.}~\bibnamefont
  {Anderson}},\ }\href {\doibase
  http://dx.doi.org/10.1016/0021-9797(89)90045-3} {\bibfield  {journal}
  {\bibinfo  {journal} {Journal of Colloid and Interface Science}\ }\textbf
  {\bibinfo {volume} {127}},\ \bibinfo {pages} {388 } (\bibinfo {year}
  {1989})}\BibitemShut {NoStop}%
\bibitem [{\citenamefont {Long}\ and\ \citenamefont
  {Ajdari}(1998)}]{AjdariLong}%
  \BibitemOpen
  \bibfield  {author} {\bibinfo {author} {\bibfnamefont {D.}~\bibnamefont
  {Long}}\ and\ \bibinfo {author} {\bibfnamefont {A.}~\bibnamefont {Ajdari}},\
  }\href {\doibase 10.1103/PhysRevLett.81.1529} {\bibfield  {journal} {\bibinfo
   {journal} {Phys. Rev. Lett.}\ }\textbf {\bibinfo {volume} {81}},\ \bibinfo
  {pages} {1529} (\bibinfo {year} {1998})}\BibitemShut {NoStop}%
\bibitem [{\citenamefont {Chae}\ and\ \citenamefont
  {Lenhoff}(1995)}]{CHAE:1995qd}%
  \BibitemOpen
  \bibfield  {author} {\bibinfo {author} {\bibfnamefont {K.~S.}\ \bibnamefont
  {Chae}}\ and\ \bibinfo {author} {\bibfnamefont {A.~M.}\ \bibnamefont
  {Lenhoff}},\ }\href {\doibase 10.1016/S0006-3495(95)80286-9} {\bibfield
  {journal} {\bibinfo  {journal} {Biophysical Journal}\ }\textbf {\bibinfo
  {volume} {68}},\ \bibinfo {pages} {1120} (\bibinfo {year}
  {1995})}\BibitemShut {NoStop}%
\bibitem [{\citenamefont {Allison}(2001)}]{Allison:2001db}%
  \BibitemOpen
  \bibfield  {author} {\bibinfo {author} {\bibfnamefont {S.}~\bibnamefont
  {Allison}},\ }\href {\doibase 10.1016/S0301-4622(01)00221-6} {\bibfield
  {journal} {\bibinfo  {journal} {BIOPHYSICAL CHEMISTRY}\ }\textbf {\bibinfo
  {volume} {93}},\ \bibinfo {pages} {197} (\bibinfo {year} {2001})}\BibitemShut
  {NoStop}%
\bibitem [{\citenamefont {Delgado}\ \emph {et~al.}(2007)\citenamefont
  {Delgado}, \citenamefont {Gonz{\'a}lez-Caballero}, \citenamefont {Hunter},
  \citenamefont {Koopal},\ and\ \citenamefont {Lyklema}}]{Delgado:2007qy}%
  \BibitemOpen
  \bibfield  {author} {\bibinfo {author} {\bibfnamefont {{\'A}.~V.}\
  \bibnamefont {Delgado}}, \bibinfo {author} {\bibfnamefont {F.}~\bibnamefont
  {Gonz{\'a}lez-Caballero}}, \bibinfo {author} {\bibfnamefont {R.}~\bibnamefont
  {Hunter}}, \bibinfo {author} {\bibfnamefont {L.}~\bibnamefont {Koopal}}, \
  and\ \bibinfo {author} {\bibfnamefont {J.}~\bibnamefont {Lyklema}},\
  }\href@noop {} {\bibfield  {journal} {\bibinfo  {journal} {Journal of colloid
  and interface science}\ }\textbf {\bibinfo {volume} {309}},\ \bibinfo {pages}
  {194} (\bibinfo {year} {2007})}\BibitemShut {NoStop}%
\bibitem [{\citenamefont {Moths}\ and\ \citenamefont
  {Witten}(2013)}]{Moths-Witten2}%
  \BibitemOpen
  \bibfield  {author} {\bibinfo {author} {\bibfnamefont {B.}~\bibnamefont
  {Moths}}\ and\ \bibinfo {author} {\bibfnamefont {T.~A.}\ \bibnamefont
  {Witten}},\ }\href {\doibase 10.1103/PhysRevE.88.022307} {\bibfield
  {journal} {\bibinfo  {journal} {Phys. Rev. E}\ }\textbf {\bibinfo {volume}
  {88}},\ \bibinfo {pages} {022307} (\bibinfo {year} {2013})}\BibitemShut
  {NoStop}%
\bibitem [{\citenamefont {Han}\ \emph {et~al.}(2016)\citenamefont {Han},
  \citenamefont {Wu},\ and\ \citenamefont {Luijten}}]{Han:2016cw}%
  \BibitemOpen
  \bibfield  {author} {\bibinfo {author} {\bibfnamefont {M.}~\bibnamefont
  {Han}}, \bibinfo {author} {\bibfnamefont {H.}~\bibnamefont {Wu}}, \ and\
  \bibinfo {author} {\bibfnamefont {E.}~\bibnamefont {Luijten}},\ }\href
  {\doibase 10.1140/epjst/e2015-50316-9} {\bibfield  {journal} {\bibinfo
  {journal} {European Physical Journal-Special Topics}\ }\textbf {\bibinfo
  {volume} {225}},\ \bibinfo {pages} {685} (\bibinfo {year}
  {2016})}\BibitemShut {NoStop}%
\bibitem [{\citenamefont {Sacanna}\ \emph {et~al.}(2011)\citenamefont
  {Sacanna}, \citenamefont {Irvine}, \citenamefont {Rossi},\ and\ \citenamefont
  {Pine}}]{Sacanna:2011fk}%
  \BibitemOpen
  \bibfield  {author} {\bibinfo {author} {\bibfnamefont {S.}~\bibnamefont
  {Sacanna}}, \bibinfo {author} {\bibfnamefont {W.~T.}\ \bibnamefont {Irvine}},
  \bibinfo {author} {\bibfnamefont {L.}~\bibnamefont {Rossi}}, \ and\ \bibinfo
  {author} {\bibfnamefont {D.~J.}\ \bibnamefont {Pine}},\ }\href@noop {}
  {\bibfield  {journal} {\bibinfo  {journal} {Soft Matter}\ }\textbf {\bibinfo
  {volume} {7}},\ \bibinfo {pages} {1631} (\bibinfo {year} {2011})}\BibitemShut
  {NoStop}%
\bibitem [{\citenamefont {Meng}\ \emph {et~al.}(2010)\citenamefont {Meng},
  \citenamefont {Arkus}, \citenamefont {Brenner},\ and\ \citenamefont
  {Manoharan}}]{Meng560}%
  \BibitemOpen
  \bibfield  {author} {\bibinfo {author} {\bibfnamefont {G.}~\bibnamefont
  {Meng}}, \bibinfo {author} {\bibfnamefont {N.}~\bibnamefont {Arkus}},
  \bibinfo {author} {\bibfnamefont {M.~P.}\ \bibnamefont {Brenner}}, \ and\
  \bibinfo {author} {\bibfnamefont {V.~N.}\ \bibnamefont {Manoharan}},\ }\href
  {\doibase 10.1126/science.1181263} {\bibfield  {journal} {\bibinfo  {journal}
  {Science}\ }\textbf {\bibinfo {volume} {327}},\ \bibinfo {pages} {560}
  (\bibinfo {year} {2010})}\BibitemShut {NoStop}%
\bibitem [{\citenamefont {Burelbach}\ and\ \citenamefont
  {Stark}(2019)}]{Burelbach:2019sz}%
  \BibitemOpen
  \bibfield  {author} {\bibinfo {author} {\bibfnamefont {J.}~\bibnamefont
  {Burelbach}}\ and\ \bibinfo {author} {\bibfnamefont {H.}~\bibnamefont
  {Stark}},\ }\href {\doibase 10.1140/epje/i2019-11769-y} {\bibfield  {journal}
  {\bibinfo  {journal} {The European Physical Journal E}\ }\textbf {\bibinfo
  {volume} {42}} (\bibinfo {year} {2019}),\
  10.1140/epje/i2019-11769-y}\BibitemShut {NoStop}%
\bibitem [{\citenamefont {Happel}\ and\ \citenamefont
  {Brenner}(1983)}]{Happel-Brenner}%
  \BibitemOpen
  \bibfield  {author} {\bibinfo {author} {\bibfnamefont {J.}~\bibnamefont
  {Happel}}\ and\ \bibinfo {author} {\bibfnamefont {H.}~\bibnamefont
  {Brenner}},\ }\href {https://books.google.com/books?id=tWO2xJZbweIC} {\emph
  {\bibinfo {title} {Low Reynolds number hydrodynamics: with special
  applications to particulate media}}},\ Mechanics of Fluids and Transport
  Processes\ (\bibinfo  {publisher} {Springer Netherlands},\ \bibinfo {year}
  {1983})\BibitemShut {NoStop}%
\bibitem [{\citenamefont {Mowitz}\ and\ \citenamefont
  {Witten}(2017)}]{Mowitz:2017kx}%
  \BibitemOpen
  \bibfield  {author} {\bibinfo {author} {\bibfnamefont {A.~J.}\ \bibnamefont
  {Mowitz}}\ and\ \bibinfo {author} {\bibfnamefont {T.~A.}\ \bibnamefont
  {Witten}},\ }\href {\doibase 10.1103/PhysRevE.96.062613} {\bibfield
  {journal} {\bibinfo  {journal} {Phys. Rev. E}\ }\textbf {\bibinfo {volume}
  {96}},\ \bibinfo {pages} {062613} (\bibinfo {year} {2017})}\BibitemShut
  {NoStop}%
\bibitem [{\citenamefont {Braverman}\ \emph {et~al.}()\citenamefont
  {Braverman}, \citenamefont {Mowitz},\ and\ \citenamefont
  {Witten}}]{Braverman:fq}%
  \BibitemOpen
  \bibfield  {author} {\bibinfo {author} {\bibfnamefont {L.}~\bibnamefont
  {Braverman}}, \bibinfo {author} {\bibfnamefont {A.~J.}\ \bibnamefont
  {Mowitz}}, \ and\ \bibinfo {author} {\bibfnamefont {T.~A.}\ \bibnamefont
  {Witten}},\ }\href@noop {} {\enquote {\bibinfo {title} {To be published},}\
  }\BibitemShut {NoStop}%
\bibitem [{Note1()}]{Note1}%
  \BibitemOpen
  \bibinfo {note} {The explicit form of $\protect \mathsf {L \kern -4pt L}$ is
  well known and is called the Oseen tensor\cite {Happel-Brenner}. However the
  present derivation doesn't depend on this explicit form. One may readily
  verify that the Oseen tensor satisfies the symmetry property shown for
  $\protect \mathsf {L \kern -4pt L}_{ij}$ below.}\BibitemShut {Stop}%
\bibitem [{Note2()}]{Note2}%
  \BibitemOpen
  \bibinfo {note} {The diagonal elements $\protect \mathsf {L \kern -4pt
  L}_{ii}$ are not defined for point stokeslets. A better approximation for the
  velocity at $\protect \mathaccentV {vec}17Er_i$ owing to the force at site
  $i$ is required. It is convenient to replace point the point force by
  distributing it uniformly over over some small region representing the
  continuum force on the fluid near site $i$. This procedure is not unique, but
  its effect on determining the $\protect \mathaccentV {vec}17Ef_i$ diminishes
  as the number of stokeslets increases. The validity of the reciprocal
  relation of Sec. \ref {sec:reciprocity} relies only on using the same
  regularization and hence the same $\protect \mathsf {L \kern -4pt L}$ for
  determining both velocity fields considered.}\BibitemShut {Stop}%
\bibitem [{Note3()}]{Note3}%
  \BibitemOpen
  \bibinfo {note} {A similar procedure treats motion caused by an external
  force. Here one requires that all the stokeslets move at a common velocity
  $\protect \mathaccentV {vec}17EV$ and that the stokeslets provide sufficient
  screening that the interior of the object also moves at velocity $\protect
  \mathaccentV {vec}17EV$. Then the total of the stokeslet forces is the force
  required for this motion\cite {Mowitz:2017kx}}\BibitemShut {NoStop}%
\bibitem [{Note4()}]{Note4}%
  \BibitemOpen
  \bibinfo {note} {The zeta potential that determines the slip velocity is the
  potential of the object surface relative to the bulk fluid. It depends on the
  local ionic environment of the surface in equilibrium without applied field
  $\protect \mathaccentV {vec}17EE_0$. It also depends on the local charge
  density on the surface and is proportional to this density for weakly charged
  regions. Typical colloids in aqueous solvents have $\zeta $ in the range of
  tens of millivolts. It is measured for a given type of surface by
  electrophoresis on a uniformly charged body with that type of
  surface.}\BibitemShut {Stop}%
\bibitem [{Note5()}]{Note5}%
  \BibitemOpen
  \bibinfo {note} {For general stokeslet objects, this locality may not be well
  defined, since an arbitrary set of stokeslets need not resemble any smooth
  surface. However, if stokeslets are arranged over a smooth surface with
  spacing much smaller than the local inverse curvature, the stokeslet object
  may approximate the corresponding smooth body, as noted above. Then the above
  reasoning applies, a stokes mobility tensor may be determined, and the
  $\protect \mathaccentV {vec}17EV$ and $\protect \mathaccentV {vec}17E\Omega $
  may be calculated\protect {\cite {Mowitz:2017kx}}}\BibitemShut {NoStop}%
\bibitem [{Note6()}]{Note6}%
  \BibitemOpen
  \bibinfo {note} {The functional form of this falloff is the well-known Oseen
  tensor mentioned above.}\BibitemShut {Stop}%
\bibitem [{\citenamefont {Goldfriend}\ \emph {et~al.}(2016)\citenamefont
  {Goldfriend}, \citenamefont {Diamant},\ and\ \citenamefont
  {Witten}}]{TomerHaim2}%
  \BibitemOpen
  \bibfield  {author} {\bibinfo {author} {\bibfnamefont {T.}~\bibnamefont
  {Goldfriend}}, \bibinfo {author} {\bibfnamefont {H.}~\bibnamefont {Diamant}},
  \ and\ \bibinfo {author} {\bibfnamefont {T.~A.}\ \bibnamefont {Witten}},\
  }\href {\doibase 10.1103/PhysRevE.93.042609} {\bibfield  {journal} {\bibinfo
  {journal} {Phys. Rev. E}\ }\textbf {\bibinfo {volume} {93}},\ \bibinfo
  {pages} {042609} (\bibinfo {year} {2016})}\BibitemShut {NoStop}%
\bibitem [{\citenamefont {Goldfriend}\ \emph {et~al.}(2015)\citenamefont
  {Goldfriend}, \citenamefont {Diamant},\ and\ \citenamefont
  {Witten}}]{TomerHaim1}%
  \BibitemOpen
  \bibfield  {author} {\bibinfo {author} {\bibfnamefont {T.}~\bibnamefont
  {Goldfriend}}, \bibinfo {author} {\bibfnamefont {H.}~\bibnamefont {Diamant}},
  \ and\ \bibinfo {author} {\bibfnamefont {T.~A.}\ \bibnamefont {Witten}},\
  }\href {\doibase 10.1063/1.4936894} {\bibfield  {journal} {\bibinfo
  {journal} {Physics of Fluids}\ }\textbf {\bibinfo {volume} {27}},\ \bibinfo
  {pages} {123303} (\bibinfo {year} {2015})}\BibitemShut {NoStop}%
\bibitem [{\citenamefont {Kirkwood}\ and\ \citenamefont
  {Riseman}(1948)}]{Kirkwood-Riseman}%
  \BibitemOpen
  \bibfield  {author} {\bibinfo {author} {\bibfnamefont {J.~G.}\ \bibnamefont
  {Kirkwood}}\ and\ \bibinfo {author} {\bibfnamefont {J.}~\bibnamefont
  {Riseman}},\ }\href {\doibase 10.1063/1.1746947} {\bibfield  {journal}
  {\bibinfo  {journal} {The Journal of Chemical Physics}\ }\textbf {\bibinfo
  {volume} {16}},\ \bibinfo {pages} {565} (\bibinfo {year} {1948})}\BibitemShut
  {NoStop}%
\bibitem [{\citenamefont {Zottl}\ and\ \citenamefont
  {Stark}(2014)}]{Zottl:2014ye}%
  \BibitemOpen
  \bibfield  {author} {\bibinfo {author} {\bibfnamefont {A.}~\bibnamefont
  {Zottl}}\ and\ \bibinfo {author} {\bibfnamefont {H.}~\bibnamefont {Stark}},\
  }\href {\doibase 10.1103/PhysRevLett.112.118101} {\bibfield  {journal}
  {\bibinfo  {journal} {Phys. Rev. Lett.}\ }\textbf {\bibinfo {volume} {112}},\
  \bibinfo {pages} {118101} (\bibinfo {year} {2014})}\BibitemShut {NoStop}%
\end{thebibliography}%

\end{document}